\documentclass[usenatbib,apjl]{emulateapj}

\newcommand{\teff}{\mbox{$T_{\rm eff}$}}
\newcommand{\logg}{\mbox{$\log g$}}
\newcommand{\vsini}{\mbox{$v \sin i$}}
\newcommand{\mictrb}{\mbox{$\xi_{\rm t}$}}
\newcommand{\mactrb}{\mbox{$v_{\rm mac}$}}

\newcommand{\kms}{\mbox{km\,s$^{-1}$}}

\newcommand{\mjup}{\mbox{M$_{\rm Jup}$}}
\newcommand{\rjup}{\mbox{R$_{\rm Jup}$}}
\newcommand{\mstar}{\mbox{$M_{*}$}}
\newcommand{\rstar}{\mbox{$R_{*}$}}
\newcommand{\msol}{\mbox{$\,{\rm M}_\odot$}}
\newcommand{\rsol}{\mbox{$\,{\rm R}_\odot$}}

\usepackage{amsmath}
\usepackage{wasysym}
\usepackage{subfigure}

\shorttitle{WASP-36 b}
\shortauthors{A. M. S. Smith et al.}

\begin{document}

\title{WASP-36\lowercase{b}: A new transiting planet around a metal-poor G-dwarf, and an investigation into analyses based on a single transit light curve}

\author{A.~M.~S.~Smith\altaffilmark{1}, D.~R.~Anderson\altaffilmark{1}, A.~Collier Cameron\altaffilmark{2}, M.~Gillon\altaffilmark{3}, C.~Hellier\altaffilmark{1}, M.~Lendl\altaffilmark{4}, P.~F.~L.~Maxted\altaffilmark{1}, D.~Queloz\altaffilmark{4}, B.~Smalley\altaffilmark{1}, A.~H.~M.~J.~Triaud\altaffilmark{4}, R.~G.~West\altaffilmark{5}, S.~C.~C.~Barros\altaffilmark{6}, E.~Jehin\altaffilmark{3}, F.~Pepe\altaffilmark{4}, D.~Pollacco\altaffilmark{6}, D.~Segransan\altaffilmark{4}, J.~Southworth\altaffilmark{1}, R.~A.~Street\altaffilmark{7}, and S.~Udry\altaffilmark{4}}

\email{amss@astro.keele.ac.uk}

\altaffiltext{1}{Astrophysics Group, Keele University, Staffordshire, ST5 5BG, UK}
\altaffiltext{2}{SUPA, School of Physics \& Astronomy, University of St Andrews, North Haugh, Fife, KY16 9SS, UK}
\altaffiltext{3}{Institut d'Astrophysique et de G\'{e}ophysique, Universit\'{e} de Li\`{e}ge, All\'{e}e du 6 Ao\^{u}t, 17, B\^{a}t. B5C, Li\`{e}ge 1, Belgium}
\altaffiltext{4}{Observatoire de Gen\`{e}ve, Universit\'{e} de Gen\`{e}ve, 51 Chemin des Maillettes, 1290 Sauverny, Switzerland}
\altaffiltext{5}{Department of Physics \& Astronomy, University of Leicester, Leicester, LE1 7RH, UK}
\altaffiltext{6}{Astrophysics Research Centre, School of Mathematics \& Physics, Queen's University, University Road, Belfast, BT7 1NN, UK}
\altaffiltext{7}{Las Cumbres Observatory, 6740 Cortona Drive Suite 102, Goleta, CA 93117, USA}

\begin{abstract}
We report the discovery, from WASP and CORALIE, of a transiting exoplanet in a 1.54-d orbit. The host star, WASP-36, is a magnitude $V=12.7$, metal-poor G2 dwarf ($\teff = 5959\pm134$ K), with [Fe/H]~$= -0.26 \pm0.10$. We determine the planet to have mass and radius respectively $2.30\pm0.07$ and $1.28\pm0.03$ times that of Jupiter.

We have eight partial or complete transit light curves, from four different observatories, which allows us to investigate the potential effects on the fitted system parameters of using only a single light curve. We find that the solutions obtained by analysing each of these light curves independently are consistent with our global fit to all the data, despite the apparent presence of correlated noise in at least two of the light curves.

\end{abstract}


\section{Introduction}

Of the 171 confirmed transiting planetary systems\footnote{http://www.exoplanet.eu, 2011 October 13}, the majority have been discovered from the ground, from surveys such as WASP \citep{Pollacco06} and HATnet \citep{HATnet}. Although the {\em Kepler} space mission is discovering an increasing number of planets and even more candidate planets (e.g. \citealt{Borucki10}; \citealt{Borucki11}), the ground-based discoveries have the advantage that the host stars are generally brighter. This allows radial-velocity measurements to measure the planetary mass, and is conducive to further characterisation observations, such as measuring occultations in the infrared to probe atmospheric temperature and structure.

Many of the current questions in exoplanet science are being addressed by analysing the statistical properties of the growing ensemble of well characterised transiting planetary systems. Here we report the discovery of a transiting planet orbiting the $V \sim 12.7$ star WASP-36 (= 2MASS J08461929-0801370) in the constellation Hydra.

\section{Observations}

\subsection{WASP photometry}

WASP-36 was observed in 2009 and 2010 by WASP-South, which is located at the South African Astronomical Observatory (SAAO), near Sutherland in South Africa, and by SuperWASP at the Observatorio del Roque de los Muchachos on La Palma, Spain. Each instrument consists of eight Canon 200mm f/1.8 lenses, each equipped with an Andor $2048\times2048$ e2v CCD camera, on a single robotic mount. Further details of the instrument, survey and data reduction procedures are described in \cite {Pollacco06} and details of the candidate selection procedure can be found in \cite{Cameron-etal07} and \cite{wasp3}. A total of 13781 measurements of WASP-36 were made between 2009 January 14 and 2010 April 21.

WASP-South 2009 data revealed the presence of a transit-like signal with a period of $\sim 1.5$ days and a depth of $\sim 15$ mmag. The WASP light curve is shown folded on the best-fitting orbital period in  Figure \ref{fig:phot}.

\subsection{Spectroscopy}

Spectroscopic observations of WASP-36 were made with the CORALIE spectrograph of the 1.2-m Euler-Swiss telescope. Simultaneous spectra of a thorium-argon emission line lamp were obtained in order to calibrate the stellar spectra. A total of nineteen spectra were taken between 2010 March 11 and 2011 January 11, and processed using the standard CORALIE data reduction pipeline \citep{ELODIE}. The resulting radial-velocity data are given in Table \ref{tab:rv_data}, and plotted in Figure \ref{fig:RV}. In order to rule out non-planetary causes for the radial-velocity variation, such as a blended eclipsing binary system, we examined the bisector spans (e.g. \citealt{Queloz01}), which exhibit no correlation with radial velocity (Figure. \ref{fig:RV}), as expected.

\begin{table} 
\caption{Radial-velocity (RV) and line bisector span (BS) measurements of WASP-36} 
\label{tab:rv_data} 
\begin{tabular*}{0.5\textwidth}{@{\extracolsep{\fill}}cccr} 
\hline 
BJD(UTC) & RV & $\sigma$$_{\rm RV}$ & BS~~~~~\\ 
 --2\,450\,000& (km s$^{-1}$) & (km s$^{-1}$) & (km s$^{-1}$)\\ 
\hline
5266.6926 & $- 12.843$ & 0.027 &$ 0.079$\\
5293.5807 & $-13.591$ & 0.021 & $-0.017$\\
5304.6409 & $-13.212$ & 0.057 & $0.025$\\
5305.5493 & $-13.423$ & 0.025 & $-0.046$\\
5306.6464 & $-12.847$ & 0.025 & $0.014$\\
5315.5600 & $-12.981$ & 0.023 & $0.013$\\
5316.5339 & $-13.624$ & 0.028 & $-0.005$\\
5317.5642 & $-12.977$ & 0.033 & $-0.046$\\
5320.4833 & $-12.821$ & 0.030 & $-0.034$\\
5359.4568 & $-13.527$ & 0.040 & $0.002$\\
5547.8347 & $-12.842$ & 0.024 & $0.007$\\
5561.8393 & $-12.866$ & 0.026 & $0.003$\\
5562.8651 & $-13.306$ & 0.025 & $-0.018$\\
5563.8184 & $-13.402$ & 0.033 & $-0.070$\\
5564.7341 & $-12.920$ & 0.024 & $-0.011$\\
5565.8211 & $-13.467$ & 0.029 & $0.056$\\
5567.8088 & $-12.870$ & 0.023 & $0.001$\\
5570.7984 & $-12.931$ & 0.023 & $-0.028$\\
5572.7427 & $-12.944$ & 0.023 & $-0.068$\\
\hline 
\end{tabular*} 
\end{table}

\subsection{Follow-up photometry}

We have a total of eight high-precision follow-up light curves of the transit of WASP-36b, summarised in Table \ref{tab:fup}. Differential aperture photometry was performed using the IRAF/DAOPHOT package for TRAPPIST and FTN data, and the ULTRACAM pipeline \citep{Ultracam, barros11} for the LT data, with aperture radii and choice of comparison stars optimised to give the lowest RMS of the out-of-transit photometry.

\begin{table*} 
\caption{Observing log for follow-up transit photometry} 
\label{tab:fup} 
\begin{tabular*}{\textwidth}{@{\extracolsep{\fill}}clcccccccc} 
\hline 
Light& Date/UT & Telescope / instrument& Band & $N_{\mathrm{obs}}$ & $t_\mathrm{exp}$ &  full or & Seeing or & Aperture & Airmass\\ 
curve &&&&&/s& partial&defocus (\arcsec)&radius (\arcsec)&range\\
\hline
(i) & 2010 Dec 13 & Euler\footnote{1.2-m Euler-Swiss Telescope, La Silla, Chile} / EulerCam & Gunn $r$ & 94&120  &partial & 1.1 -- 2.2& 4.3&1.54 -- 1.07 -- 1.09\\
(ii) & 2010 Dec 13 & TRAPPIST\footnote{Transiting Planets and Planetesimals Small Telescope, La Silla, Chile (\citealt{TRAPPIST}; http://www.astro.ulg.ac.be/Sci/Trappist)}  / TRAPPISTCAM & {\em clear} & 756& 10  & partial & 3 & 8.3 & 1.84 -- 1.11\\
(iii) & 2010 Dec 25 & FTN\footnote{Faulkes Telescope North, Haleakala Observatory, Maui, Hawaii, USA} / Spectral camera& PS $z$ & 176&60 & full & 4.3 & 2.4 & 1.52 -- 1.14 -- 1.22\\
(iv) & 2011 Jan 02 & TRAPPIST / TRAPPISTCAM & $I + z$ & 296& 25 & full & 2 &6.4&1.75 -- 1.09 \\
(v) & 2011 Jan 05 & TRAPPIST / TRAPPISTCAM & \em{clear} & 179& 18  &partial & 3 &7.7 & 1.18 -- 1.07 -- 1.14\\
(vi) & 2011 Jan 08 & TRAPPIST / TRAPPISTCAM & \em{clear} & 269& 18  & partial & 3.5&9.0&1.10 -- 1.44\\
(vii) & 2011 Jan 15/16 & LT\footnote{Liverpool Telescope, Observatorio del Roque de los Muchachos, La Palma, Spain} / RISE\footnote{Rapid Imaging Search for Exoplanets camera (\citealt{rise2008, gibson2008})} & $V + R$ & 1290& 9 &  full & 6&9.0 & 1.95 -- 1.25\\
(viii) & 2011 Jan 21 & Euler / EulerCam & Gunn $r$ & 167 & 60  &full & 0.45 -- 1.0 & 4.1&1.46 -- 1.20\\
\hline 
\end{tabular*} 
\end{table*} 

\section{Determination of system parameters}

\subsection{Stellar parameters}

The individual CORALIE spectra of WASP-36 were co-added to produce a
single spectrum with a typical S/N of around 60:1. The standard CORALIE pipeline
reduction products were used in the analysis. The spectral analysis was performed using {\sc uclsyn} \citep{uclsyn1,uclsyn2} and the methods given in
\citet{2009A&A...496..259G}. The parameters obtained from
the analysis are listed in Table~\ref{tab:stellar_params}. The elemental abundances
were determined from equivalent width measurements of several clean and
unblended lines. The lines used are those listed in \cite{lines1}, \cite{lines2}, and \cite{lines3}. A value for microturbulence, \mictrb, was determined from
Fe~{\sc i} using the method of \cite{1984A&A...134..189M}. The quoted error
estimates account for the uncertainties in \teff, \logg\ and \mictrb,
as well as for the scatter due to measurement and atomic data uncertainties.

The projected stellar rotation velocity, \vsini, was determined by fitting the
profiles of several unblended Fe~{\sc i} lines in the wavelength range 6000--6200\AA, using the rotation broadening function of \cite[Ch. 18]{gray-book}. We used an instrumental FWHM of 0.11 $\pm$ 0.01 \AA, determined from the telluric lines around 630 nm. The measured \vsini\ is sensitive to the adopted value of \mactrb. The appropriate value of \mactrb is 4.0 \kms\ according to \cite[p. 507]{gray-book}, but 2.9 \kms\ according to \cite{2010MNRAS.405.1907B}. These values imply \vsini\ = 2.9 $\pm$ 1.3~\kms\ and \vsini\ = 3.7 $\pm$ 1.1~\kms\, respectively. We take the weighted average of these two values as the best-fitting one, \vsini\ = 3.3 $\pm$ 1.2~\kms\ . The quantity measured is approximately the quadratic sum of \vsini\ and \mactrb\ ($\approx 4.9 \pm 0.8$~\kms\ ).

\begin{table}[h]
\caption{Stellar parameters and abundances from analysis of CORALIE spectra.}
\begin{tabular}{ll|cc} \hline
Parameter  & Value & Parameter & Value \\ \hline
RA (J2000.0) & 08h46m19.30s & {[Fe/H]}   & $-$0.26 $\pm$ 0.10 \\
Dec (J2000.0) & $-08^{\circ}~01\arcmin~36.7\arcsec $& {[Na/H]}   & $-$0.33 $\pm$ 0.08 \\
\teff      &  5900 $\pm$ 150 K & {[Mg/H]}   & $-$0.08 $\pm$ 0.08 \\
\logg (cgs)     & 4.5 $\pm$  0.15 & {[Si/H]}   &$-$0.17 $\pm$  0.06 \\
\mictrb\footnote{Microturbulent velocity}    &  1.0  $\pm$ 0.2 \kms & {[Ca/H]}   & $-$0.15 $\pm$ 0.11 \\
\vsini     &  3.3 $\pm$ 1.2 \kms & {[Sc/H]}   & $-$0.11 $\pm$ 0.12 \\
log A(Li)\footnote{$\mathrm{log~A(Li)} = \log\left(\frac{N_\mathrm{Li}}{N_\mathrm{H}}\right) + 12$, where $N_\mathrm{Li}$ and $N_\mathrm{H}$ are the number densities of Li and H respectively.}  &    1.69 $\pm$ 0.13 & {[Ti/H]}   & $-$0.16 $\pm$ 0.11 \\
Sp. Type   &   G2 & {[V/H]}    & $-$0.20 $\pm$ 0.15 \\
Distance   &    450 $\pm$ 120 pc & {[Cr/H]}   &$-$0.28 $\pm$ 0.09 \\
Age & 1 -- 5 Gy & {[Mn/H]}   &$-$0.44 $\pm$ 0.10 \\
Mass       &    1.01 $\pm$ 0.08 $\msol$&{[Co/H]}   &$-$0.19 $\pm$ 0.12 \\
Radius     &   0.94 $\pm$ 0.17 $\rsol$ &{[Ni/H]}   &$-$0.30 $\pm$ 0.08 \\
\hline
\multicolumn{4}{l}{Additional identifiers for WASP-36:}\\
\multicolumn{4}{l}{USNO-B1.0 0819-0221838}\\
\multicolumn{4}{l}{2MASS J08461929-0801370}\\
\multicolumn{4}{l}{1SWASP J084619.30-080136.7}\\
\hline
\\
\end{tabular}
\label{tab:stellar_params}
\newline {\bf Note:} The spectral type was estimated from \teff\
using the table of \cite[p. 507]{gray-book}. The mass and radius were estimated using the \cite{2010A&ARv..18...67T} calibration.
\end{table}

\subsection{Neighbouring objects}

The Two Micron All Sky Survey catalogue \citep{2MASS} reveals the presence of four fainter stars close on the sky to WASP-36. There is no evidence from analysis of catalogue proper motions that any of these stars are physically associated with WASP-36. The stars are separated from WASP-36 by 4\arcsec, 9\arcsec, 13\arcsec and 17\arcsec, meaning that they fall well within the WASP photometric aperture, which has a radius of 48\arcsec (3.5 pixels), but outside of the 1\arcsec~CORALIE fibre.

In the absence of reliable optical catalogue magnitudes for all of these objects, it was necessary to measure their fluxes to quantify the effects of blending in the photometry. The fluxes were measured from images taken during the two transits observed with the 1.2-m Euler-Swiss Telescope (see Table \ref{tab:fup}). The fluxes relative to that of WASP-36 are as follows: 0.012 (object at 4\arcsec~separation from WASP-36), 0.00771 (9\arcsec), 0.00558 (13\arcsec) and 0.00827 (17\arcsec). Using these flux ratios, we corrected the WASP photometry to account for all four objects, and the high precision photometry to account for the object at 4\arcsec, which is within the photometric apertures used. The magnitude of this correction is minimal, and had no significant ($\ll$~1-$\sigma$) effect on the values of our best-fitting system parameters.

\subsection{Planetary system parameters}
\label{sec:mcmc}

CORALIE radial-velocity data were combined with all our photometry and analysed simultaneously using the Markov Chain Monte Carlo (MCMC) method.  The best-fitting system parameters are taken to be the median values of the posterior probability distribution. Linear functions of time were fitted to each light curve at each step of the MCMC, to remove systematic trends.  We use the current version of the MCMC code described in \cite{Cameron-etal07, wasp3} and \cite{Enoch10}. The MCMC proposal parameters we use are: the epoch of mid-transit, $T_{\rm c}$; the orbital period, $P$; the transit duration, $T_{\rm 14}$; the fractional flux deficit that would be observed during transit in the absence of stellar limb-darkening, $\Delta F$; the transit impact parameter, $b$; the stellar reflex velocity semi-amplitude, $K_{\mathrm{1}}$; the stellar effective temperature, $T_{\mathrm{eff}}$; the stellar metallicity, [Fe/H]; and $\sqrt{e}\cos\omega$ and $\sqrt{e}\sin\omega$, where $e$ is the orbital eccentricity, and $\omega$ is the argument of periastron \citep{wasp30}. The stellar mass was determined as part of the MCMC analysis, using an empirical fit to [Fe/H], \teff, and the stellar density, $\rho_{\rm *}$ (\citealt{Enoch10}; \citealt{2010A&ARv..18...67T}).

The transit light curves were modelled using the formulation of \cite{M&A} and limb-darkening was accounted for using a four-coefficient, non-linear model, employing coefficients appropriate to the passband from the tabulations of \cite{Claret, Claret04}. The coefficients were determined using an initial interpolation in $\log g_{*}$ and [Fe/H] (values from Table \ref{tab:stellar_params}), and an interpolation in $T_{\rm *, eff}$  at each MCMC step. The coefficient values corresponding to the best-fitting value of $T_{\rm *, eff}$ are given in Table \ref{tab:limb}. Because some of our photometry was observed in passbands not tabulated by  \cite{Claret, Claret04}, we also tried using coefficients corresponding to nearby passbands. None of our best-fitting system parameters was significantly affected by our choice of Claret passband; values changed by much less than their 1-$\sigma$ uncertainties.

An initial MCMC fit for an eccentric orbit found $e = 0.012^{+ 0.014}_{- 0.008}$ ($\omega = 55^{+55}_{-118}$ degrees), with a 3-$\sigma$ upper-limit to the eccentricity of 0.064, but we found this eccentricity is not significant. Following the {\em F}-test approach of \cite{lucy_sweeney}, we find that there is a 58 per cent probability that the apparent eccentricity could have arisen if the underlying orbit were actually circular. We therefore present here the model with a circular orbit, noting that the values of the other model parameters, and their associated uncertainties, are almost identical to those of the eccentric solution. 

We tried fitting for a linear trend in the RVs with the inclusion of an additional parameter in our MCMC fit. Such a trend (such as that found in the RVs of WASP-34, \citealt{wasp34}) would be indicative of a third body in the system. The best-fitting radial acceleration is consistent with zero, indicating there is no evidence for an additional body in the system based on our RVs, which span 10 months. The orbital parameters we report are the result of a fit which does not allow for a linear trend in radial velocity.

\begin{figure} 
\includegraphics[width=21 cm,angle=270]{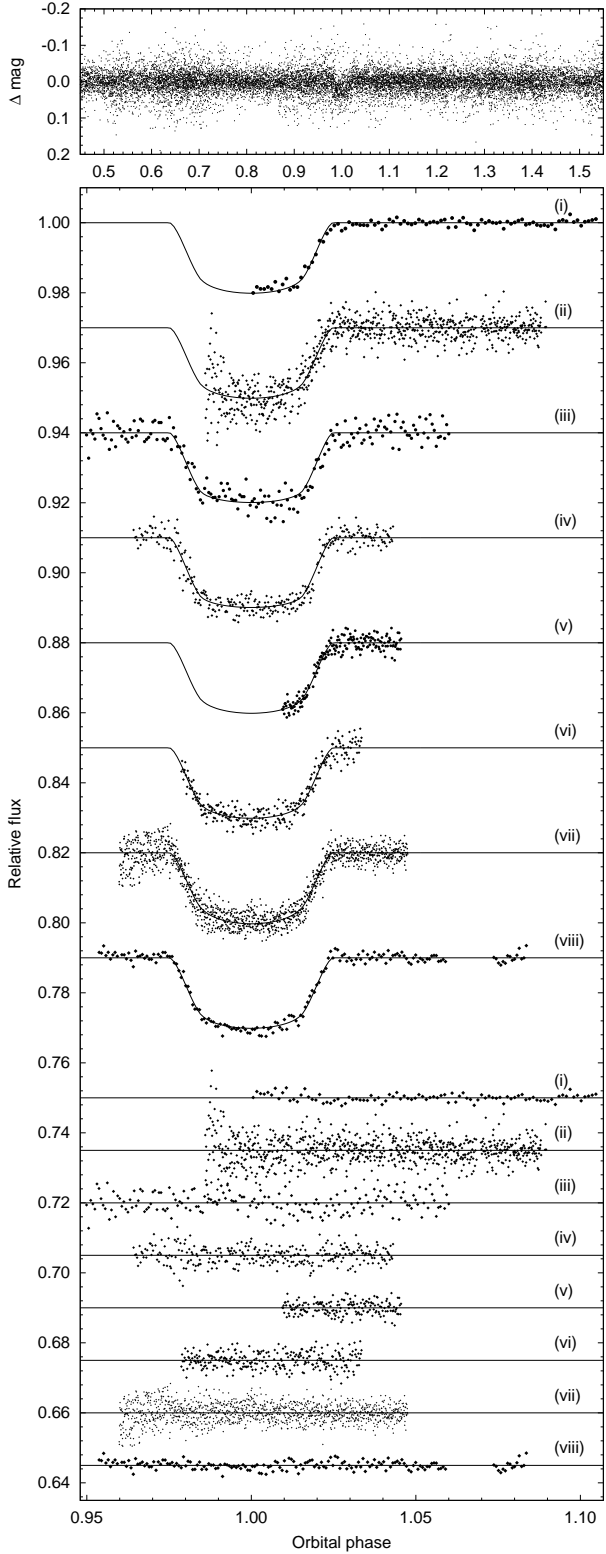} 
\caption{Photometry. {\it Upper panel:} WASP-36 b discovery light curve folded on the orbital period of $P = 1.5373653~d$. For display purposes, points with an error greater than three times the median uncertainty are not shown. 
{\it Lower panel:} High-precision transit photometry, over-plotted with our best-fitting model (solid lines). Each individual dataset is offset in flux for clarity, and is labelled with a numeral corresponding to that in the first column of Table \ref{tab:fup}. The residuals to the best-fitting models are labelled and offset in the same way, and are shown in the lower part of the panel.}
\label{fig:phot} 
\end{figure}

\begin{figure} 
\includegraphics[width=0.7\textwidth,angle=270]{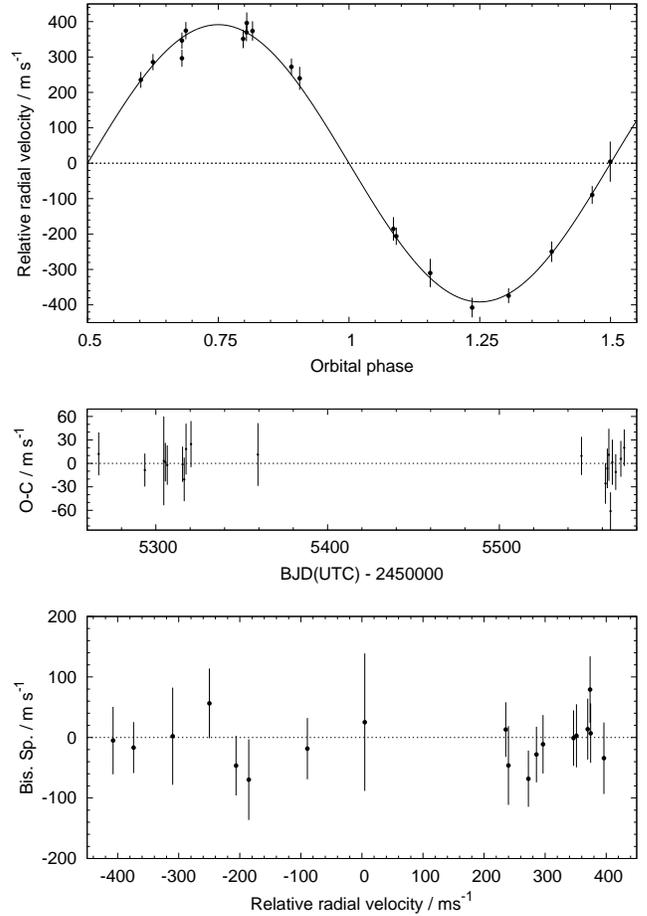} 
\caption{Radial velocities. {\it Upper panel:} Phase-folded radial-velocity measurements (Table \ref{tab:rv_data}. The centre-of-mass velocity, $\gamma$ = -13.2169 km s$^{-1}$, has been subtracted. The best-fitting MCMC solution is over-plotted as a solid line. {\it Middle panel:} Residuals from the radial-velocity fit as a function of time. {\it Lower panel:} Bisector span measurements as a function of radial velocity. The uncertainties in the bisectors are taken to be twice the uncertainty in the radial velocities.
}
\label{fig:RV} 
\end{figure}

The system parameters derived from our best-fitting circular model are presented in Table \ref{tab:sys_params}. The corresponding transit and RV models are superimposed on our data in Figures \ref{fig:phot} and \ref{fig:RV}.

\begin{table}
\label{tab:limb}
\centering
\caption{Limb-darkening coefficients}
\begin{tabular}{cccccc} \hline 
Claret band & Light curves & $a_1$ &  $a_2$  &  $a_3$ & $a_4$  \\
\hline 
Cousins $R$  & WASP,i,ii,v,vi,viii & 0.466 &  0.294  & 0.070  &-0.128 \\
Sloan $z^\prime$         & iii,iv	    	                   & 0.555 & -0.099  & 0.348  &-0.213 \\
Johnson $V$ & vii 	                   & 0.389 &  0.523  &-0.066  &-0.082 \\
\hline 
\end{tabular} 
\end{table} 

\begin{table*} 
\caption{System parameters} 
\label{tab:sys_params} 
\begin{tabular*}{\textwidth}{@{\extracolsep{\fill}}lccc} 
\hline 
Parameter & Symbol & Unit & Value \\ 
\hline 
\\
Orbital period	    	    	    	    & 	$P$ & d & 1.5373653$\pm{0.0000026}$\\
Epoch of mid-transit	    	    	    & 	$T_{\rm c}$ &HJD, UTC & 2455569.83731$\pm{0.000095}$\\
Transit duration    	    	    	    & 	$T_{\rm 14}$ &d & 0.07566$\pm0.00042$\\
Ingress / egress duration   	    	    & 	$T_{\rm 12}=T_{\rm 34}$ &d & 0.01540$\pm{0.00054}$\\
Planet-to-star area ratio   	    	    & 	$\Delta F=R_{\rm P}^{2}$/R$_{*}^{2}$&- & 0.01916$\pm{0.00020}$\\
Transit impact parameter    	    	    & 	$b$ &-& 0.665$\pm{0.013}$\\
Orbital inclination angle   	    	    & 	$i$ &$^\circ$  & 83.61$\pm{0.21}$\\
Stellar orbital velocity semi-amplitude     & 	$K_{\rm *}$ &km s$^{-1}$ & 0.3915$\pm{0.0083}$\\
System velocity     	    	    	    &     	$\gamma$ &km s$^{-1}$ &$ -13.2169 \pm{0.0024}$\\
Orbital eccentricity (adopted)	    	    	    & 	$e$ &-& 0\\
Orbital eccentricity (3-$\sigma$ upper-limit) & &-& $0.0663$\\
Stellar mass	    	    	    	    & 	$M_{\rm *}$ & $M_{\rm \odot}$ & 1.040$\pm{0.031}$\\
Stellar radius	    	    	    	    & 	$R_{\rm *}$ & $R_{\rm \odot}$ & 0.951$\pm{0.018}$\\
log (stellar surface gravity)     	    	    & 	$\log g_{*}$ & (cgs) & 4.499$\pm{0.012}$\\
Stellar density     	    	    	    & 	$\rho_{\rm *}$ &$\rho_{\rm \odot}$ & 1.211$\pm0.050$\\
Stellar effective temperature         &  $T_{\rm *, eff}$ & K & 5959$\pm{134}$\\
Metallicity					    &  [Fe/H] & dex & $-0.26\pm{0.10}$\\
Planet mass 	    	    	    	    & 	$M_{\rm P}$ &$M_{\rm Jup}$ & 2.303$\pm{0.068}$\\
Planet radius	    	    	    	    & 	$R_{\rm P}$ &$R_{\rm Jup}$ & 1.281$\pm{0.029}$\\
Planet density	    	    	    	    & 	$\rho_{\rm P}$ &$\rho_{\rm J}$ & 1.096$\pm{0.067}$\\
log (planet surface gravity)     	    	    & 	$\log g_{\rm P}$ & (cgs) & 3.507$\pm{0.018}$\\
Scaled orbital major semi-axis     &   $a/R_{\rm *}$ &-& 5.977$\pm${0.082}\\
Orbital major semi-axis     	    	    & 	$a$ &AU  & 0.02643$\pm{0.00026}$\\
Planet equilibrium temperature (uniform heat redistribution)	    	    & 	$T_{\rm P, A=0}$ &K & 1724$\pm{43}$\\
System age (from figure \ref{fig:hr}) &&Gy&$2.5^{+3.5}_{-2.2}$\\ 
\footnotetext[0]{The following constant values are used: AU $= 1.49598\times10^{11}$~m; $R_{\rm \odot} = 6.9599\times10^8$~m; $M_{\rm \odot} = 1.9892\times10^{30}$~kg; $R_{\rm Jup} = 7.1492\times10^7$~m; $M_{\rm Jup} = 1.89896\times10^{27}$~kg; $\rho_{\rm J} = 1240.67$~kg m$^{-3}$}. \\ 
\hline 

\end{tabular*} 
\end{table*}

\begin{figure} 
\includegraphics[width=0.45\textwidth]{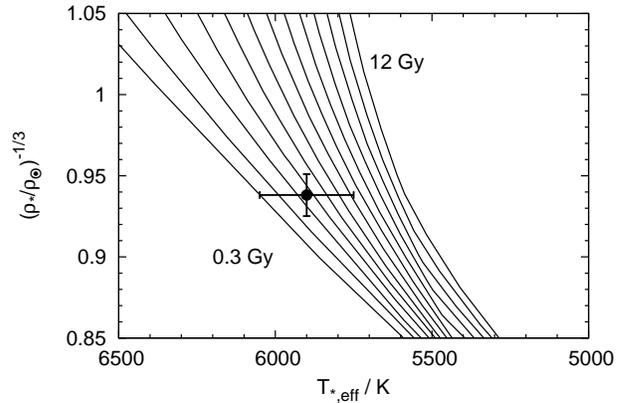} 
\caption{Modified Hertzsprung-Russell diagram. WASP-36 is plotted alongside isochrones from the evolutionary models of \cite{Marigo}. The isochrones are for $Z = 0.01044$ ([Fe/H] $= -0.26$). The leftmost curve corresponds to an age of 0.3 Gy, and the adjacent curve to 1.0 Gy. Thereafter the curves are regularly spaced at intervals of 1~Gy.
}
\label{fig:hr} 
\end{figure}

\begin{figure} 
\includegraphics[width=9cm,angle=0]{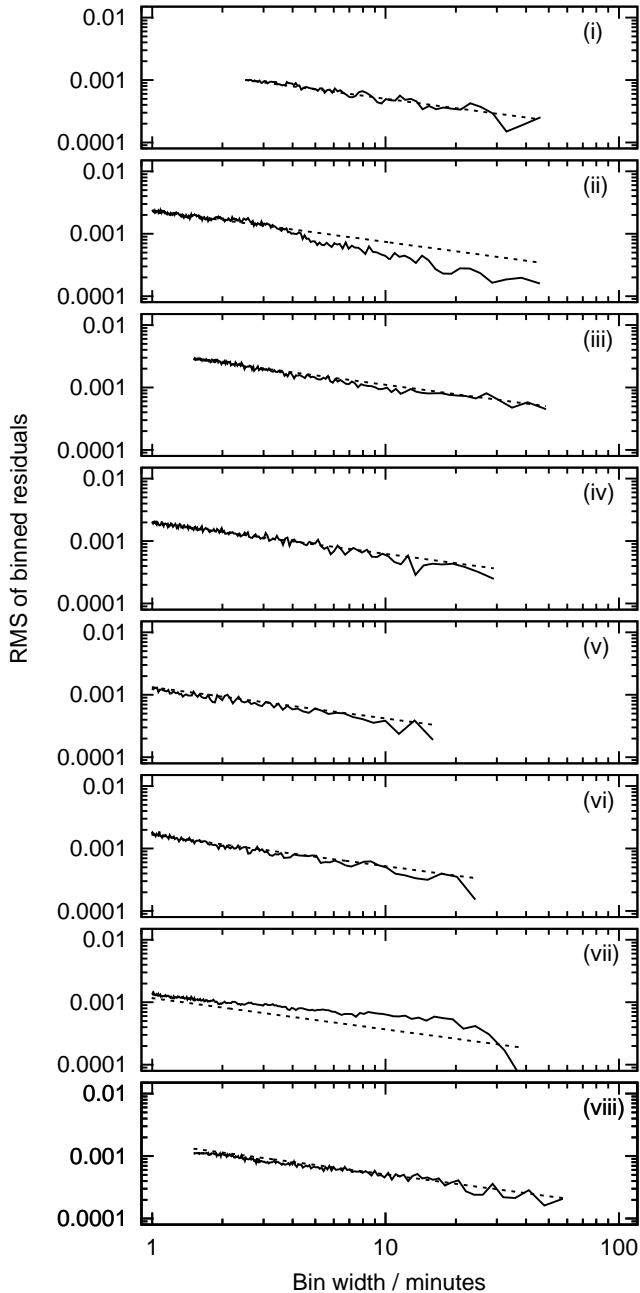} 
\caption{Correlated noise in follow-up light curves -- RMS of binned residuals versus bin width, for light curves i - viii (solid lines). The white-noise expectation, where the RMS decreases in proportion to the square root of the bin size, is indicated by the straight, dotted lines. }
\label{fig:rms} 
\end{figure} 

\begin{table}
\label{tab:ttv}
\centering
\caption{Transit times}
\begin{tabular}{crccr} \hline 
Light curve & $E$ &  $T_C$  &  $\sigma_{T_C}$ & $O-C$  \\
& & (HJD, UTC) & (min) & (min) \\
\hline 
(i) & $-17$  &  2455543.70602 &  5.72  &  5.65  \\
(ii) & $-17$  &  2455543.70378 &  1.18  &  2.42  \\
(iii) & $-9$  &  2455556.00221 &  0.62  &  1.71  \\
(iv) &  $-4$ &  2455563.68807 &  0.33  &  0.32  \\
(v) &  $-2$ &  2455566.76718 &  7.73  &  6.63  \\
(vi) & $0$  &  2455569.83686 &  0.86  &  $-0.62$  \\
(vii) &  $5$ &  2455577.52412 &  0.21  &  $-0.02$  \\
(viii) &  $9$ &  2455583.67344 &  0.27  &  $-0.23$  \\
\hline 
\end{tabular} 
\end{table} 

\begin{figure*} 
\includegraphics[width=16cm,angle=0]{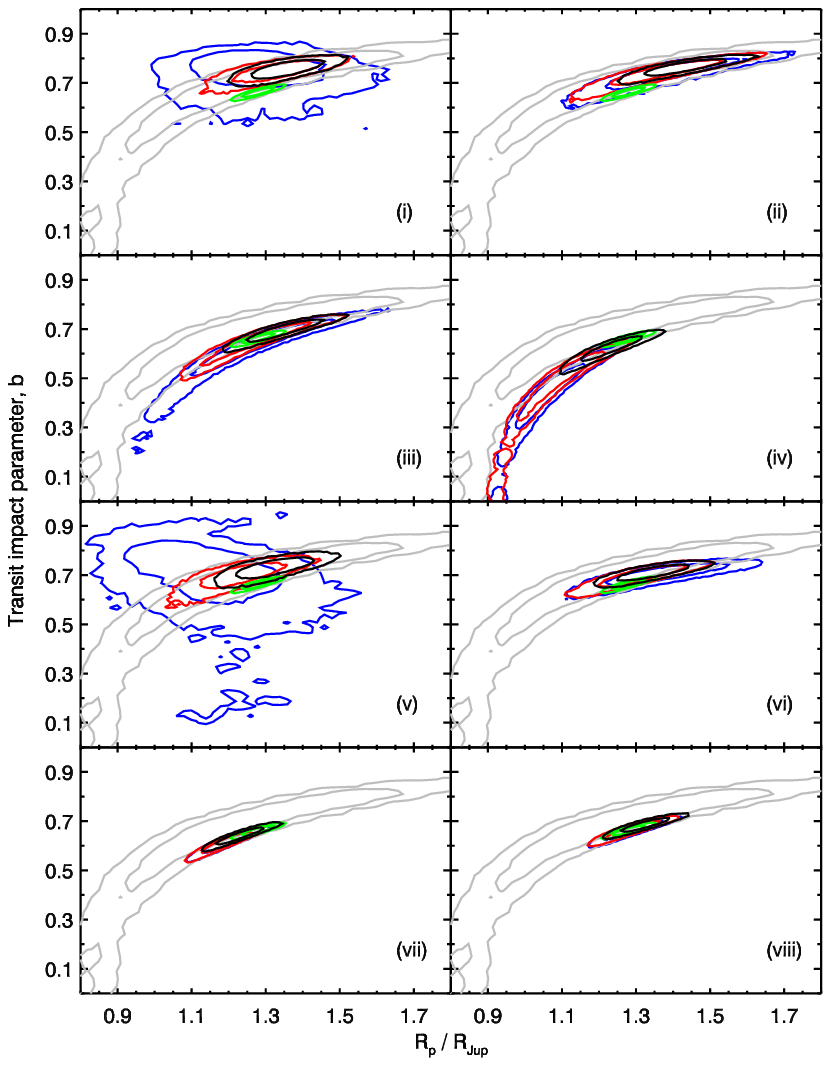} 
\centering
\caption{Analysis of follow-up light curves I. The MCMC posterior probability distributions for $R_\mathrm{P}$ and $b$ for each of the follow-up light curves. The numbering of each panel corresponds to the light curve numbering in Table \ref{tab:fup} and the 1-$\sigma$ and 2-$\sigma$ contours are shown. In each case red corresponds to analysis of a single follow-up light curve plus the WASP photometry, black to the single light curve plus the WASP photometry with the main-sequence constraint imposed, and blue to that of a single light curve with {\em no} WASP photometry. The green contours indicate our global solution, and the grey contours the solution without follow-up photometry}, and are therefore identical in each panel.
\label{fig:bRp} 
\end{figure*} 

\begin{figure*} 
\includegraphics[width=16cm,angle=0]{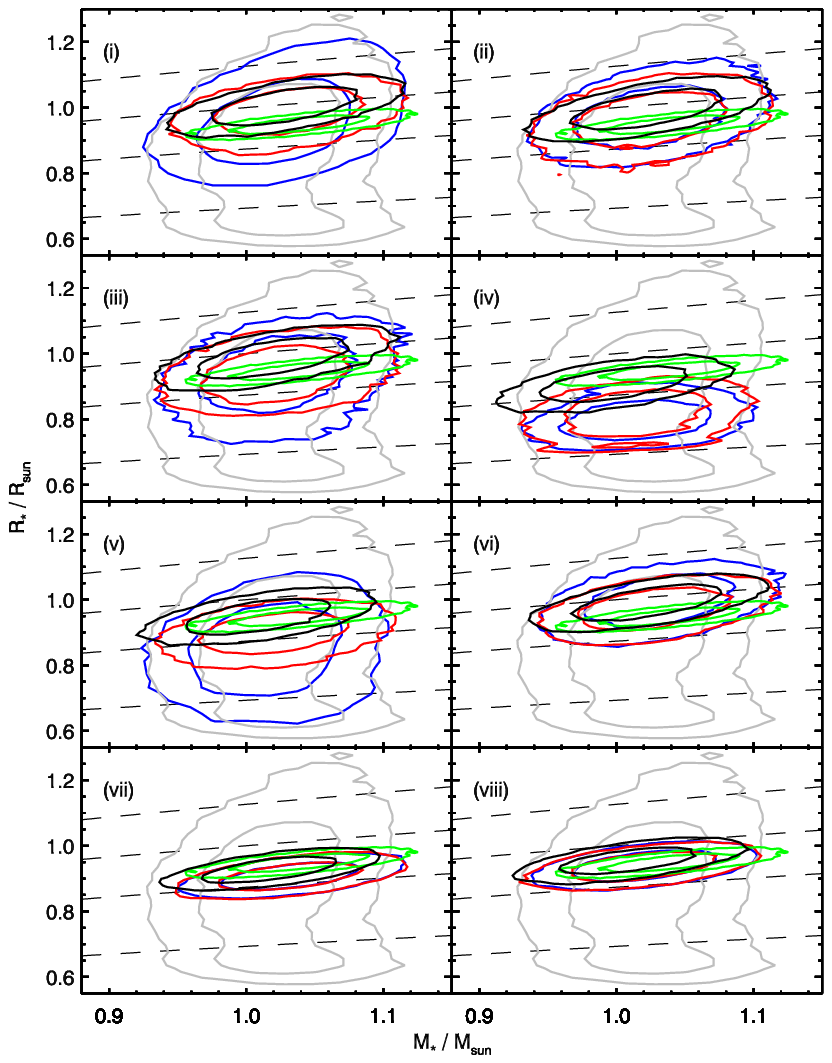} 
\centering
\caption{Analysis of follow-up light curves II. The MCMC posterior probability distributions for $\mstar$ and $\rstar$ for each of the follow-up light curves. The numbering of each panel corresponds to the light curve numbering in Table \ref{tab:fup} and the 1-$\sigma$ and 2-$\sigma$ contours are shown. In each case red corresponds to analysis of a single follow-up light curve plus the WASP photometry, black to the single light curve plus the WASP photometry with the main-sequence constraint imposed, and blue to that of a single light curve with {\em no} WASP photometry. The green contours indicate our global solution, and the grey contours the solution without follow-up photometry, and are therefore identical in each panel. Also in each panel are dashed lines which are contours of constant stellar density, corresponding, from top to bottom, to 0.7, 1.0, 1.5, and 3.0 times solar density.}
\label{fig:RsMs} 
\end{figure*} 

\subsection{System age}

The measured $v \sin i$ of WASP-36 gives an upper limit to the rotational period, $P_{\rm rot} \simeq 14.4 \pm 5.9$~days. This corresponds to an upper limit on the age of $\sim~1.8^{+2.7}_{-1.3}$~Gy using the gyrochronological relation of
\citet{2007ApJ...669.1167B}, and a $B$-magnitude of 13.3 derived from $V$=12.7 and $B-V = 0.60 \pm 0.04$  (estimated using \citealt[p. 507]{gray-book}).

In Figure \ref{fig:hr} we plot WASP-36 alongside the stellar evolution tracks of \cite{Marigo}. From this we infer an age of $2.5^{+ 3.5}_{- 2.2}$ ~Gy. The age determined from the lithium abundance of WASP-36 is poorly constrained, but the work of \cite{2005A&A...442..615S} suggests that the most likely age is $\sim~2$ to 5 Gy.

We searched the WASP photometry for periodic variations indicative of star spots and stellar rotation, but no significant variation was detected. We place an upper limit of 1.5 mmag at the 95 per cent confidence level on the amplitude of any sinusoidal variation. This null result is consistent with the low levels of stellar activity expected from a main-sequence G2 star. A lack of stellar activity is also indicated by the absence of calcium II H+K emission in the spectra. The uncertainties on the CaII emission index, $\log R^\prime _{HK} (\approx 4.5)$ are too large to allow meaningful constraints to be placed on the system age by using an activity - rotation relation such as that of \cite{mamajek_hillenbrand}.

There is no evidence of any discrepancy between the ages derived from lithium abundance, gyrochronology and isochrone fitting. This suggests that the star has undergone little or no tidal spin-up, despite the presence of a massive planet in a close orbit.

\subsection{Transit timing}

We measured the times of mid-transit for each of the eight follow-up light curves, by analysing each light curve separately, without any other photometry (see Section \ref{sec:8lcmethod}). The times are displayed in Table \ref{tab:ttv}, along with the differences, $O - C$ between these times and those predicted assuming a fixed epoch and period (Table \ref{tab:sys_params}). No significant departure from a fixed ephemeris is observed.

\section{Detailed analaysis of follow-up light curves}

Because we have several follow-up light curves of WASP-36 from different telescopes / instruments, whereas many planet discovery papers rely on only a single such light curve, we take the opportunity here to examine in detail the potential effects on the system parameters of using only a single light curve.

For survey photometry with low SNR, the durations of ingress and egress are ill-defined, leading to considerable uncertainty in the transit impact parameter and hence to large uncertainties in the stellar density and planetary radius. So-called `follow-up' transit light curves are generally included in the analysis of new ground-based transiting planet discoveries, and are of significantly higher photometric precision than the light curves produced by survey instruments such as WASP. Such follow-up light curves are typically the result of observations with a 1--2-m class telescope, and are of critical importance to measuring precisely basic system parameters.

Any light curve may suffer from correlated noise, such as from observational systematics or from astrophysical sources such as stellar activity. To assess the levels of correlated noise in our follow-up light curves, we plot (Figure \ref{fig:rms}) the RMS of the binned residuals to the fit of each light curve as a function of bin width, along with the white-noise expectation. For six of our light curves, the RMS of the binned residuals follows closely the white-noise expectation, indicating that little or no correlated noise is present in the data. Light curves ii and vii show deviation from the white-noise model, however, suggesting the presence of noise correlated on timescales of $\sim 1$ and $\sim 10$ minutes, respectively. We suggest that this red noise may be due to the high airmass of the target at the start of these observations.

\subsection{Method}
\label{sec:8lcmethod}
After modelling all available data in a combined MCMC analysis (see Section \ref{sec:mcmc}), our `global solution', we also ran several MCMCs each with just a single follow-up light curve in addition to the radial velocities and WASP photometry. Additionally, we re-ran each of these MCMCs applying a Gaussian prior to the stellar radius to impose a density typical of a main-sequence star (the `main-sequence constraint'). Such a constraint is usually applied when analysing a new planet which has poor quality follow-up photometry (such as a single light curve which covers only part of the transit), and there is no evidence that the star is evolved or otherwise non-main-sequence in nature. We also performed analyses where the only photometry included was a single follow-up light curve, i.e. the WASP photometry was excluded from the analysis. The purpose of this is to determine whether the measured depth of transit is biased by inclusion of the WASP photometry. For these runs only, the orbital period was fixed to the value determined as part of our global solution, since this parameter  is very poorly constrained by a single transit light curve and a few RVs. The epoch of mid-transit was treated as normal, and allowed to float freely. Finally, we performed an analysis exlcuding all follow-up photometry; the only photometry analysed was the WASP data.

\subsection{Results}

We produced correlation plots between several parameters, but choose to present here only plots showing impact parameter against planet radius and stellar radius versus stellar mass (Figures \ref{fig:bRp} and \ref{fig:RsMs}, respectively). Such plots, whilst representative of the ensemble correlation plots, are particularly instructive since $b$ and $R_\mathrm{P}$ are two of the major quantities we wish to measure, are largely constrained by follow-up light curves rather than by survey photometry or by radial velocities, and can be significantly correlated with each other, indicating a strongly degenerate solution. The stellar density is measured directly from the transit light curve, and the stellar mass and radius, whilst interesting in themselves, are key in determining the values of several other system parameters of interest.

Several conclusions can be drawn from study of Figures \ref{fig:bRp} and \ref{fig:RsMs}, and similar plots, namely:

(1) each analysis including only a single follow-up light curve gives results that are consistent with our global solution, albeit with larger uncertainties. To measure the dispersion in the best-fitting parameter values obtained from each single follow-up light curve analysis, we calculated the weighted standard deviation. The standard deviations of $b$, $R_\mathrm{P}$, $\rstar$ and $\mstar$ are 0.05, 0.08 $\rjup$, 0.05 $R_{\rm \odot}$ and 0.006 $M_{\rm \odot}$, respectively.

(2) The largest uncertainties are obtained for follow-up light curves that cover the smallest fraction of the transit (light curves i and v), as expected. 

(3) The analyses which exclude the WASP photometry give larger uncertainties, but these are only significantly so when the follow-up photometry is poor. This indicates that the WASP photometry only makes a significant contribution to constraining the shape and depth of the transit when the follow-up light curve is incomplete. 

(4) Even a partial transit light curve improves the precision of the measured system parameters enormously compared to those derived solely from the WASP photometry and the RVs.

(5) The imposition of a main-sequence constraint does not significantly alter the parameters or uncertainties for high-precision light curves that are complete, thus indicating that WASP-36 is a main-sequence star. When the follow-up light curve does not well constrain the range of possible models, however, limiting the star to the main-sequence can significantly reduce the large degeneracy in the possible solutions. This is best illustrated by light curve (iv), where the effects of the constraint are to decrease the stellar density we find and confine the solution to a smaller area of parameter-space, close to the global solution, while largely resolving the degeneracy between $b$ and $R_\mathrm{P}$.

In summary, if only one of the follow-up light curves had been available, we would have reached a solution compatible with the current best-fitting model, although the uncertainties on the model parameters may have been much greater, if the light curve was not of the highest precision. Obtaining additional light curves is clearly of benefit if one only has a light curve that partially covers transit. It is also useful to have multiple high-precision light curves for systems where stellar activity may bias the observed transit depth by varying amounts at different epochs, as may be the case for WASP-10b \citep{wasp10,johnson_wasp10,dittmann_wasp10,maciejewski_wasp10_i,maciejewski_wasp10_ii}.

\section{Discussion and conclusion}

WASP-36 is a metal-poor, Solar-mass star which is host to a transiting planet in a 1.54~d orbit. We find the planet to have a mass of 2.30 \mjup, and a radius 1.28 \rjup, meaning it is slightly denser than Jupiter. There is an observed correlation between planetary radius and insolation (e.g. \citealt{wasp25}), with the more bloated planets generally receiving a greater flux from their star. WASP-36b is somewhat larger than predicted by the models of \cite{BLL}, which predict radii between 1.08 \rjup (for a planet with a core at 1500 K) and 1.20 \rjup (for a core-less planet at 2000 K).

The close orbit and large radius of the planet make it a good target for measuring the planetary thermal emission, via infra-red secondary eclipse (occultation) measurements with, for example, {\it Spitzer}. The expected signal-to-noise ratios of the occultations in {\it Spitzer} channels 1 (3.6 $\mu$m) and 2 (4.5 $\mu$m) are around 10 and 9 respectively, assuming a planet with zero albedo and uniform heat redistribution.

One of the striking properties of the WASP-36 is the low stellar metallicity ([Fe/H] = $-0.26\pm0.10$). Giant planets are known to be rare around such low-metallicity stars (e.g. \citealt{santos04,F&V05}), although several other low-metallicity systems are known, including the transiting systems WASP-21 ([Fe/H] = $-0.46\pm0.11$, \citealt{wasp21}), WASP-37 ([Fe/H] = $-0.40\pm0.12$, \citealt{wasp37}) and HAT-P-12 ([Fe/H] = $-0.29\pm0.05$, \citealt{hat12}).

Such systems will be critical in probing our understanding of the planet--metallicity correlation; proposed explanations for the correlation include insufficient material for proto-planetary cores to attain the critical mass needed for runaway accretion, and the suggestion that the high density of molecular hydrogen in the inner galactic disk is responsible for the effect \citep{haywood09}. WASP-36b may also play a key role in determining whether stellar metallicity is the key parameter influencing whether or not a hot Jupiter's atmosphere exhibits a thermal inversion. Insolation was initially propounded as this parameter \citep{fortney}, but several planets now appear to contradict this theory. XO-1b, for instance, has a relatively low insolation, and was therefore predicted to lack an inversion, but \cite{Machalek08} report the presence of an inversion; TrES-3b does not exhibit an inversion \citep{Fressin10} despite a prediction to the contrary. More recently stellar activity \citep{knutson_stellar_activity} and stellar metallicity \citep{wheatley_metallicity} have been advanced as alternatives to insolation; work aiming to resolve this issue is ongoing.

\section{Acknowledgments}

WASP-South is hosted by the SAAO and SuperWASP by the Isaac Newton Group and the Instituto de Astrof\'{i}sica de Canarias;  we gratefully acknowledge their ongoing support and assistance. Funding for WASP comes from consortium universities and from the UK's Science and Technology Facilities Council (STFC).  TRAPPIST is a project funded by the Belgian Fund for Scientific Research (FNRS) with the participation of the Swiss National Science Fundation. MG and EJ are FNRS Research Associates. The RISE instrument mounted in the Liverpool Telescope was designed and built with resources made available from Queen's University Belfast, Liverpool John Moores University and the University of Manchester.  The Liverpool Telescope is operated on the island of La Palma by Liverpool John Moores University in the Spanish Observatorio del Roque de los Muchachos of the Instituto de Astrof\'{i}sica de Canarias with financial support from the STFC. We thank Tom Marsh for use of the ULTRACAM pipeline. We thank the anonymous referee for their thorough comments which improved the final manuscript.

\bibliography{iau_journals,refs2}
\bibliographystyle{apj}

\end{document}